# Einstein and the mistery of eternity of life.

by

M.Barone
**Institute of Nuclear Physics**
NCSR "Demokritos"
Athens-Greece
e-mail : barone@inp.demokritos.gr

and

M.Gajewska
**Institute of Philosophy and Sociology**
University of Gdansk-Poland
e-mail : gajagajewska@wp.pl

## Abstract

While Special and General Relativity are well known by a very large number of people, Einstein's religious convinctions are almost ignored.
The paper focuses on what was his view about immortality.Einstein believed in Spinoza's view of God and in cosmic religion.Since the Special Relativity implies a kind of superdeterminism with the future completely determined,that can be taken as ontological proof for the non-Existence of an hypothetical God.
Because of that K.Popper in a private conversation with him said: "You are Parmenides". Einstein's Pantheism was in contrast with the non-Euclidean universe of Special Relativity.
In conclusion the interest in our times of his Weltanschaung will be taken in consideration.

# Einsteins 's death.

Einstein died at the age of seventy-six of a burst aneurysm in Princeton,New Jersey,on April 18,1955.In the summer of 1950,his doctors found that a weak blood vessel on his abdominal aorta was getting larger.The doctors wrapped the inflamed blood vessel with a cellophane to avoid a hemorage and treated him with drugs,since Einstein refused any additional surgery intervention to correct the problem.On March 18,1950 he signed his will in which he dictated his funeral details; he wanted a simple ceremony and no gravestone. His brain was removed and set aside of his body and according to his whishes and contrary to the Jewish tradition he asked to be cremated and have his ashes scattered.
The way Einstein left the human life make us understand what was his opinion about the death and immortality.We will review here in his ideas about this matter.

# Religion

Einstein, as German law required, had an official religious education.His family,fairly unobservant Jewish, hired a distant relative as tutor and he at the age of eleven he embraced Judaism with ferocity,throwing himself into Jewish traditions,including refusing to eat pork.
This phase did not last long ;at the age of twelve Einstein discovered the world of science and the bible stories now sounded like lies told to children.Therefore, he reversed his previous religiouness and for the rest of his life he never participated in a traditional religious ritual.
Starting in the late 1920s,Einstein wrote several essays on religiosity,summarized
in the book :" The World as I See it" , published in 1950s, where he speaks about a religion that offers a union with science.
In the evolution of religion in his view there were 3 stages.
At the first step he put the "religion of fear"

*"With primitive man it is above all fear that evokes religious notions-fear of hunger,wild beasts,sickness,death.Since at this stage of exhistence understanding of causal connexions is poorly developed,the human mind creates for itself more or less analogous beings on whose wills and actions these fearful happenings dependes.One's object now is to secure the favour of these beings by carrying out actions and offering sacrifices which,according to the tradition handed down from generation,propiziate them or make them well disposed toward a mortal.I am speaking now of the religion of fear.This ,though not created,is in an important degree stabilized by the formation of a special priestly caste, which set up as mediator between the people and the beings fear,and erects a hegemony on this basis.In many cases the leader or ruler whose position depends on other factors,or a privileged class,combines priesltly functions with*

*secular authority in order to make the latter more secure; or the political rulers and the priestly caste make common cause in their own interests."*

The second step is a "moral religion":

*"The social feelings are antother source of the crystallization of religion. Fathers and mothers and leaders of larger human communities are mortal and fallible. The desire for guidance, love, and support prompts men to form the social or moral conception of God.*
*This is the God of Providence who protects, disposes, rewards, and punishes, the God who, according to the with of the believer's outlook, loves and cheriches the life of the tribe or the human race................"*
*"The Jewish scriptures admirably illustrate the development from the religion of fear to moral religion, which is continued in the New Testament. The religion of all civilized peoples, especially the peoples of the Orient, are primarly moral religions. The development from the religion of fear to moral religion is a great step in a nation's life.*
*That primitives religions are based entirely on fear and the religions of civilized peoples purely on morality is a prejudice against which we must be on our guard. The truth is that they are all intermediates types, with the reservation, that on higher levels of social life the religion of morality predominates."*

Both kind of religions are characterized by the anthropomorphic caracter of their conception of God.

The third step represents a religion which is reached by exceptionally high minded people and communities: "the cosmic religion". In fact he says:

*"But there is a third state of religious experience.........., even though it is rarely found in pure form, and which I will call cosmic religious feeling. It is very difficult to explain this feeling to anyone who is entirely without it, especially as there is no anthropomorphic conception of God corresponding to it."*
*"The religious geniuses of all ages have been distinguished by this kind of religious feeling, which knows no dogma and no God conceived in man's image; so that there can be no Church whose central teaching are based on it. Hence it is precisely among the erectics of every age that we find men who were filled with the*

highest kind of religious feeling and were in many cases regarded by their contemporaries as Atheists, sometimes also as saints. Looked at in this light, men like Democritus, Francis of Assisi and Spinoza are closely akin to one another.

How can cosmic religious feeling be communicated from one person to another, if it can give rise to no definite notion of God and no theology? In my view, it is the most important function of art and science to awaken this feeling and keep it alive in those who are capable of it.

We thus arrive at a conception of the relation of science to religion very different from the usual one. When one views the matter historically one is inclined to look upon science and religion as irreconcilable antagonists, and for a very obvious reason. The man who is thoroughly convinced of the universal operation of the law of causality cannot for a moment entertain the idea of a being who interferes in the course of events - that is, if he takes the hypothesis of causality really seriously. He has no use for the religion of fear and equal little for social or moral religion. A God who rewards and punishes is inconceivable to him for the simple reason that a man's actions are determined by necessity, any more than an inanimate object is responsible for the motions it goes through. Hence science has been charged with undeterming morality, but the charge is unjust. A man's ethical behaviour should be based effectually on sympathy. education and social ties; no religious basis is necessary. Man would indeed be in poor way if he had to be restrained by fear and punishement and hope of reward after death.

It is therefore easy to see why the Churches have always fought science and persecuted its devotees. On the other hand, I maintain that cosmic religious feeing is the strongest and noblest incitement to scientific research. Only those who realize the immense effort and, above all, the devotion which pioneer work in theoretical science demands, can grasp the strength of the emotion out of which alone such work, remote as it is from the immediate realities of life, can issue. What a deep conviction of the rationality of the universe and what a yearning to understand, were it but a feble reflection of the mind revealed in this world, Kepler and Newton must have had to enable them to spend years of solitary labour in disentangling the principles of celestial mechanics! Those whose acquaintance with scientific research is derived chiefly from its pratical results easily develop a completely false notion of the mentality of the men who, surrounded by a skeptical world, have shown the way to those like-minded with themselves, scattered trough the earth and the centuries."

Einstein presented for the first time the three stages of the religious evolution in a November 9,1930 article he wrote for the New York Times Magazine called Religion and Science .It was one of his first public declarations of how he viewed science and religion. This article was met with a mixed response and he was surprised by how many people wished to discuss his ideas on religion.
The strongly conservative side derided him ,in fact a catholic priest said that Einstein had made a mistake by including the "s" in his "cosmic religion".Some liberal Jewish rabbis applauded him.

## Spinoza's view of God and religion.

Baruch Spinoza(1632-1677) was a Dutch philosopher whom Einstein claimed always as his favorite philosopher .
Einstein seems to have studied and discuss the Spinoza's Ethics(Etica Ordinae Geometrico Demonstrata),when he was in Bern.Soon after getting his first job at the patent office,he formed with M.Besso and M. Solovine a discussion circle ,called Akademia Olympia.,where he read the book.The Ethics is a book organized on the Euclidean model based on deductions from postulates,inferences,lemmas .For Spinoza ,God and Nature were one(*Deus sive Natura*).True religion was not based on dogma but on a feeling for the rationality and the unity underlying all finite and temporal things,on a feeling of surprise and fear that generates the idea of God which lacks any anthropomorphic conception.
The world is contructed on cause and effect,nothing happens that cannot be explained by the previous chain of events.
Because of his independence of mind,his deterministic philosophical outlook,his skepticism about organized religion and orthodoxy he was excommunicated from his synagogue in Amsterdam in 1656.
Einstein's cosmic religion is largely inspired by the Spinoza's view of God.
In fact the New York's Rabbi Herbert S.Goldstein ,alarmed by the suspicion of atheism appearing in the Einstein statements,asked to great physicst by telegram:

*"Do you believe in God?Stop.Answer paid 50 words"*

He replied in twenty nine words he replied

*"I believe in Spinoza's God who reveals himself in the orderly harmony of what exists,not in a God who concerns himself with fates and actions of human beings".*

Nevertheless Einstein himself was occasionally contradictory on this subject.In an interview for the book"Glimpses of the Great", published in 1930 by George Sylvester Viereck,asked specifically if he belived in the God of Spinoza,he said:

*"I can't answer with simple yes or no.I'm not an atheist and I don't think I can call myself a pantheist…..I am fascinated by Spinoza's pantheism,but admire even more his contributions to modern thought because he is the first philosopher to deal with the soul and body as one,not two separate things".*

Indeed, Spinoza's work and the way of life had been important to Einstein:he had written an introduction to a biography of Spinoza(by his son in law Rudolf Kayser) in 1946,in 1951 he had contributed to the Spinoza Dictionary and he had referred to Spinoza in many of his letters.In particular Michele Besso in a letter to Einstein written on January 19,1955 in a discussion about God, recognizes that the great physicist believes in the Spinoza's God.
Einstein's trust on causality led him to question not only the existence of soul and God who interfered with the human life,but also the physics of his days based on Quantum Mechanics whom's Bohr's interpretation was formulated on probabilityThe Universe was fundamentally random.Einstein in a discussion with the Danish physicist is reported to have said to him:

*"God does not play dice with the Universe"*

and Bohr replied

*"Quit telling God what to do"*.

Since Spinoza identified God with Nature (*Deus sive Natura*) and Einstein believed into Spinoza's God,we can conclude that Einstein was also a pantheist,because Pantheism is the religious belief in the divinity of Nature and that we humans are part of the One,interconned whole.It is realizing our connection to the One Universe above we find truth,spiritual fulfillment and solace.

### Einstein-Parmenides

Einstein in a letter sent on march 21,1955 to both the son and the sister of M.Besso with the occasion of the death of his very good friend , he said:

*"He has preceded me a little by parting from this strange world.This means nothing.To us belivieng physicists the distinction between past,present and future has only the significance of a stubborn illusion".*

Therefore since there is not distinction between,past ,present and future,then time is an illusion.

In fact, the Special Relativity, understood by Einstein as a four-dimensional space-time continuum, implies a kind of super-determinism with the future completely determined to the smallest detail.This was the reason why in the book "Unended quest,an intellectual biography"pusblished in Glasgow in 1976 at page 129,Karl Popper criticising the Einstein's belief that time is an illusion, he told him:

*"You are Parmenides"*.

For the greek philosopher(515-445 B.C) the being is not becoming and time(becoming) an illusion.
If everything is exactly pre-determined then ,there is no free will,not even a hypotetical God and a God without free will is an ontological impossibility.
We should note that if we accept the idea that there is no distinction between past,present and future, then it shoud be possible to move from the future to the past. Then some logical paradoxes appear to be in contradiction with the Special Relativity which allows only phenomena moving with a speed not exceeding that of light.The trip from future to past can be made only if we admit the existence of super-luminal signals.
Parmenides believed everything must exist,which meant to him that change was an optical illusion,therefore the existence and being is a unity.
Existence could not be created and was indestructible.

Spinoza in Ethics(Corollary 2 of Proposition 20 ) says:

*"It follows that God is immutable or,wich is the same thing, i.e.all His attributes are immutable"*.

In agreement with both philosophers,Einstein in 1917 in his attempt to apply General Relativity,to cosmology,being convinced that the universe was static and closed, he introduced the cosmological constant known as Lambda.By carefully choosing the amount of matter in the universe and the value of the cosmological constant,he could balance the gravitational forces to obtain a static universe.Some years later ,when Hubble concluded from his astronomical data that the universe was expanding,he acknowledged that the cosmological constant was his"greatest blunder".

## Immortality.

In the book "All the Questions You Ever Wanted to Ask American Atheists" by Madalyn Murray O'Hair,vol II,pag.26,1982 Einstein is reported to have said:

*"Immortality?There are two kinds.The first lives in the imagination of the people ,and is thus an illusion.There is relative immortality which may conserve the memory of an individual for some generations.But, there is only one true*

*immortality,on a cosmic scale,and that is the immortality of the cosmos itself.There is no other."*

In a letter sent to his friend M.Besso on august 10,1954,he recognized that inspite of his efforts to describe nature via Special Relativity and General Relativity,his theories could in future be wrong.Then, none of his theories could survive a longer time.
Therefore, his decision to be cremated and to have the ashes scattered in the air can be interpreted as a continuation of his life in a different form in the cosmos and forever.

J.Ph.Pierron in the book"L'Avenir de la mort,Etudes sur la mort,pag,73-83,no.121,2002 points out that cremation can be taken as a positive event because the human body does not undergo to organic decomposition ,but it is reduced to ashes,made up by small particles,"atomes",recalling to a scientist the eternity of life.

## Conclusion

The Enstein's belief is still of interest in our times :

The common man in the day life keeps away from himself the idea of death,hoping that it can arrive to other people but not to him and he is shocked when he is hit.It is the culture that allows one to go beyond such a limit by means of rites,myths, religion , rationality and science.In this way man is free from fear and aims to immortality.
The sociologist E.Morin in his book:L'homme et la mort",Paris Edition Seuil,1970, states that death could be more acceptable if immortality is taken as a prolongation of life for an indefinite period.In fact in this case death is just an event of a long and well definte route,which does not have an infinite length because in this case it implys an infinite time which brings to something divine.Instead ,a definite time implies a chrological succession of events limited to a range going from some years to hundred years.
The notion of time plays an important role. In fact, A.Barreau in his book:"Mort a' jouer mort a' dejouer socio-antopologie du mal de mort,Paris,Presses Universitaires de France,1994 , supports the thesis according to which modern man can understand the death process only in terms of the notion of time.This means that we need a definition of time.

Einstein's Weltanschaung based on his concept of time,pantheism,belief in rationality,determinism and cosmic religion,democracy and pacifism, is very close to todays aspirations of humanity.Finally, it seems that his decision to be cremated has incited people all over the world to follow his example.

Acknowledgments

Many thanks are due to Prof.A.Theophilou for fruitful discussions and to prof.L.Kostro for the stimulation given to both the authors.

## References.

1) Morin E. :  « L'homme et la mort » Paris, Edition Seuil,1970.

2) L. V. Thomas, preface of Barrau A. :  « Mort à jouer mort, mort à dèjouer socio-anthropologie du mal de mort »,Paris, Presses Universitaires de France,1994.

3) Barrau A., « Mort à jouer mort à dèjouer socio-anthropologie du mal de mort », Paris, Presses Universitaires de France, 1994.

4) Einstein A. : « Comme je vois le monde », Paris ; Flammarion,1979.English version :The world as I see it. http://www.lib.ru/FILOSOF/EJNSHTEJN/theworld_engl.txt

5) Merleau-Point J.: « Einstein », Paris, Flammarion,1993

6) Babel H. : « Dieu dans l'univers d'Einstein », Naef/RAMSAY, Genève, 2005.

7) Einstein A. : « Correspondance avec Michele Besso », Paris, Hermann, 1979

8) http://philoctetes.free.fr/parmenide.pdf

9) Holton G."Einsten's Third Paradise".Daedalus Fall 2003.

10) Winterberg F.:"The Einstein-myth and the crisis in modern physics" .Proc. of Physical Interpretations of Relativity Theory-IX,Sept.3-6,2004 Imperial College,London,UK.

11) K.C.Fox and A.Keik:"Einstein A to Z",2004 John Wiley&Sons,New Jersey,USA.